\documentclass[twocolumn,showpacs,preprintnumbers,amsmath,amssymb]{revtex4}

\usepackage{graphicx}
\usepackage{dcolumn}
\usepackage{bm}

\begin{document}


\title{Inflationary Attractor in Braneworld Scenario}

\author{Zong-Kuan Guo}
 \email{guozk@itp.ac.cn}
\author{Hong-Sheng Zhang}%
\affiliation{%
Institute of Theoretical Physics, Chinese Academy of Sciences,
P.O. Box 2735, Beijing 100080, China
}%

\author{Yuan-Zhong Zhang}
\affiliation{
CCAST (World Lab.), P.O. Box 8730, Beijing 100080\\
Institute of Theoretical Physics, Chinese Academy of Sciences,
P.O. Box 2735, Beijing 100080, China
}%

\date{\today}

\begin{abstract}
We demonstrate the attractor behavior of inflation driven by a scalar
field or a tachyon field in the context of recently proposed four-dimensional
effective gravity induced on the world-volume of a three-brane in
five-dimensional Einstein gravity, and we obtain a set of exact inflationary
solutions. Phase portraits indicate that an initial kinetic term decays rapidly
and it does not prevent the onset of inflation. The trajectories more rapidly
reach the slow rolling curve in braneworld scenario than in the standard
cosmology.
\end{abstract}

\pacs{98.80.Cq, 04.50.+h}
\maketitle

\section{Introduction}

Recent developments in string theory and its extension M-theory have suggested
that the standard model particles are confined on a hypersurface (called brane)
embedded in a higher dimensional space (called bulk). Only gravity and other
exotic matter such as the dilaton can propagate in the bulk~\cite{HW}. Our
universe may be such a brane-like object. In the braneworld scenario,
constraints on the size of extra dimensions become weaker because the standard
model particles propagate only in three spatial dimensions. Arkani-Hamed,
Dimopoulos and Dvali considered a flat bulk geometry in (4+d)-dimensions,
in which $d$ dimensions are compact with radius $R$
(toroidal topology)~\cite{ADD}. Considerable progress was made by Randall and
Sundrum, who considered non-flat, i.e. warped bulk geometries~\cite{RS}.
In their first paper, they proposed a mechanism to solve the hierarchy problem
by a small extra dimension. In their second paper, the braneworld with a
positive tension was investigated. The effective Einstein equations and the
generalized Friedmann equation on the brane were derived~\cite{SMS,BDE}.
Since the energy density of the brane appears quadratically in the new
Friedmann equation in contrast with the linear behaviour of the usual equation,
the scenario has interesting cosmological implications, in particular, the
prospects of inflation are enhanced on the brane~\cite{MWBH}.
Modifications to the Friedmann constraint equation lead to a faster Hubble
expansion at high energies and a more strongly damped evolution of the
scalar field. This assists slow rolling inflation, enhances the amount of inflation
obtained in any given model, and drives the perturbations towards an exactly
scale-invariant Harrison-Zel'dovich spectrum~\cite{MWBH}. 
Recently, several authors have investigated the cosmological implications of a
tachyon rolling down its potential, which have not been derived from string
theory, to its ground state~\cite{SEN,GWG}. Sufficient inflation driven by the
rolling tachyon can be achieved in the context of the braneworld
scenario~\cite{BBS}.

To date a question which has not yet been addressed in the literature on
inflation in the braneworld scenario is the issue of constraints on the phase
space of initial conditions for inflation. If inflation is to be truly predictive,
the evolution when the scalar field is at some given point on the potential
has to be independent of the initial conditions. Otherwise, any result, such as
the amplitude of density perturbations, would depend on the unknowable initial
conditions. However, the scalar wave equation is a second-order equation,
implying that $\dot{\phi}$, in principle, can take on any value anywhere
on the potential, and so, there certainly is not a unique solution at each point
on the potential. Therefore inflation can be predictive only if the solutions
exhibit an attractor behavior, where the differences between solutions of
different initial conditions rapidly vanish. In the standard cosmology
the inflationary equations do indeed possess this vital property~\cite{GOLD}.
In the braneworld scenario, it is interesting and necessary to investigate the
initial conditions for inflation. In this paper, using the Hamilton-Jacobi
formalism, we show that inflation driven by a scalar field or a tachyon field
on the brane possesses the attractor behavior and obtain a set of exact
inflationary solutions. As examples we plot the trajectories in the phase space
numerically. Phase portraits indicate that an initial kinetic term decays rapidly
and it does not prevent the onset of inflation. Moreover, we find that the
trajectories more rapidly reach the slow rolling curve in the braneworld
scenario than in the standard cosmology.

\section{Modified Friedmann Equation in the Braneworld Scenario}

In the five-dimensional braneworld scenario~\cite{RS}, the four-dimensional
Einstein equations induced on the brane can be written as~\cite{SMS}
\begin{equation}
G_{\mu \nu}=-\Lambda_4 g_{\mu \nu}+\kappa_4^2T_{\mu \nu}+\kappa_5^4
\pi_{\mu \nu}-E_{\mu \nu},
\end{equation}
where $T_{\mu \nu}$ is the energy-momentum tensor of matter on the
brane, $\pi_{\mu \nu}$ is a tensor quadratic in $T_{\mu \nu}$,
\begin{equation}
\pi_{\mu \nu}=\frac{1}{12}TT_{\mu \nu}-\frac{1}{4}T_{\mu \eta}T_{\nu}^{\eta}
+\frac{1}{8}g_{\mu \nu}T_{\eta \lambda}T^{\eta \lambda}-\frac{1}{24}T^2
g_{\mu \nu},
\end{equation}
and $E_{\mu \nu}$ is a projection of the five-dimensional Weyl tensor,
describing the effect of bulk graviton degrees of freedom on brane dynamics.
The effective cosmological constant $\Lambda_4$ on the brane is determined
by the five-dimensional bulk cosmological constant $\Lambda_5$ and the
brane tension $\sigma$ as
\begin{equation}
\Lambda_4=\frac{\kappa_5^2}{2}\left(\Lambda_5+\frac{\kappa_5^2}{6}
\sigma^2\right),
\end{equation}
and the four-dimensional Planck scale is given by
\begin{equation}
\kappa_4^2=\frac{\kappa_5^2}{6}\sigma.
\end{equation}
In a spatially flat Friedmann-Roberson-Walker (FRW) model the Friedmann
equation on the brane becomes~\cite{BDE}:
\begin{equation}\label{FRW}
H^2=\frac{\Lambda_4}{3}+\frac{\kappa_4^2}{3}\rho\left(1+\frac{\rho}{2\sigma}
\right)+\frac{\mu}{a^4}.
\end{equation}
where $\mu$ is an integration constant arising from $E_{\mu \nu}$, and
thus transmitting bulk graviton influence onto the brane. This term is called
the dark radiation term, which can be obtained from a full analysis of the
bulk equations. In the following we will assume that $\mu=0$ and
$\Lambda_4=0$, which indicates that these is a fine-tuning
$\Lambda_5=-\frac{\kappa_5^2}{6}\sigma^2$. If the brane tension $\sigma$
is much below the energy scale on the brane, then the new term in
$\rho^2$ is dominant in the modified Friedmann equation. Hence, the term
in $\rho$ is negligible during inflation in early universe.

\section{Attractor Behavior for Standard Inflaton}

We consider a scalar field $\phi$ confined on the brane with self-interaction
potential $V(\phi)$. The evolution equation of the field is
\begin{equation}\label{EE}
\ddot{\phi }+3H\dot{\phi }+V^\prime (\phi)=0,
\end{equation}
subject to the modified Friedmann constraint 
\begin{equation}\label{FE}
H^2=\frac{\kappa_4^2}{6\sigma}\left(\frac{1}{2}\dot{\phi}^2
 +V(\phi)\right)^2.
\end{equation}
One can derive a very useful alternative form of these equations by using the
scalar field as a time variable~\cite{LPB}. This requires that $\phi$ does not
change sign during inflation. Without loss of generality, we can choose
$\dot{\phi}>0$ throughout. If this is not satisfied, it can be brought about by
redefining $\phi \to -\phi$. Differentiating Eq.(\ref{FE}) with respect to $t$
and using Eq.(\ref{EE}) gives
\begin{equation}
\dot{H}=-\frac{3\kappa_4}{\sqrt{6\sigma}}H\dot{\phi}^2.
\end{equation}
We may divide each side by $\dot{\phi}$ to eliminate the time dependence in
the Friedmann equation, obtaining
\begin{equation}\label{HJ1}
\dot{\phi}=-\frac{\sqrt{6\sigma}}{3\kappa_4}\frac{H^\prime(\phi)}{H(\phi)},
\end{equation}
which gives the relation between $\phi$ and $t$. This allows us to write the
Friedmann equation in the first-order form
\begin{equation}\label{HJ2}
\frac{\sigma}{3\kappa_4^2}\left[\frac{H^\prime(\phi)}{H(\phi)}\right]^2-
\frac{\sqrt{6\sigma}}{\kappa_4}H(\phi)=-V(\phi).
\end{equation}
This new set of equations (\ref{HJ1}) and (\ref{HJ2}) is the Hamilton-Jacobi
equations, which is normally more convenient than the Eqs.(\ref{EE}) and
(\ref{FE}). It allows us to consider $H(\phi)$, rather than $V(\phi)$, as the
fundamental quantity to be specified. Once $H(\phi)$ has been specified, we
immediately can obtain the corresponding potential from Eq.(\ref{HJ2}). Also, 
Eq.(\ref{HJ1}) gives the relation between $\phi$ and $t$, which enables us
to obtain $H(t)$, which, if it is desired, can be integrated to $a(t)$. Therefore,
this is the most direct route to obtaining a large set of exact inflationary
solutions.

As an example of using (\ref{HJ1}) and (\ref{HJ2}), let us consider a universe
in the braneworld scenario with $H(\phi)=\phi^{-\alpha}$. Elementary algebra
now gives the potential to be of the form
\begin{equation}
V(\phi)=\frac{\sqrt{6\sigma}}{\kappa_4}\phi^{-\alpha}-\frac{\sigma \alpha^2}
{3\kappa_4^2}\phi^{-2}.
\end{equation}
The corresponding evolution of $\phi(t)$ is
\begin{equation}
\phi(t)=\left(\phi_0^2+\frac{2\sqrt{6\sigma}\alpha}{3\kappa_4}t\right)^{\frac{1}{2}}.
\end{equation}
Then we obtain
\begin{equation}
a(t) \propto \left\{ \begin{array}{ll}
 \left(1+\frac{4\sqrt{6\sigma}}{3\phi_0^2\kappa_4}t\right)^{\frac{3\kappa_4}
 {4\sqrt{6\sigma}}} & \alpha=2 \\
 \exp\left[\frac{3\kappa_4}{\alpha(2-\alpha)\sqrt{6\sigma}}\left(\phi_0^2+
 \frac{2\sqrt{6\sigma}\alpha}{3\kappa_4}t\right)^{1-\frac{\alpha}{2}}\right]
 & \alpha \ne 2.
 \end{array} \right.
\end{equation}

Using the Hamilton-Jacobi equations (\ref{HJ1}) and (\ref{HJ2}), we demonstrate
the attractor behavior in the braneworld scenario. Suppose $H_0(\phi)$ is any
solution to Eq.(\ref{HJ2}), which can be either inflationary or non-inflationary.
Add to this a linear homogeneous perturbation $\delta H(\phi)$; the attractor
condition will be satisfied if it becomes small as $\phi$ increases. For simplicity,
we shall assume, discuss further below, that the perturbations do not reverse
the sign of $\dot{\phi}$, though the result holds under more general
circumstances. Substituting $H(\phi)=H_0(\phi)+\delta H(\phi)$ into Eq.(\ref{HJ2})
and linearizing it, we find that the perturbation obeys
\begin{equation}
H_0^\prime \delta H^\prime \simeq \left(\frac{9\sqrt{6}\kappa_4}{2\sqrt{\sigma}}
H_0-\frac{3\kappa_4^2}{\sigma}V\right)H_0\delta H,
\end{equation}
which has the general solution
\begin{equation}
\delta H(\phi)=\delta H(\phi_i)\exp \left[ \int_{\phi_i}^\phi \left( \frac{3\sqrt{6}
\kappa_4}{2\sqrt{\sigma}}\frac{H_0^2}{H_0^\prime}+\frac{H_0^\prime}{H_0}\right)
d\phi\right],
\end{equation}
where $\delta H(\phi_i)$ is the value at some initial point $\phi_i$. We have
used the fact that $H_0$ is any solution to Eq.(\ref{HJ2}). Because
$H^\prime _0$ and $d\phi$ have opposing signs, the integrand within the
exponential term is negative definite, and hence all linear perturbations
do indeed die away. If there is an inflationary solution, all linear perturbations
approach it at least exponentially fast as the scalar field rolls.

To study an explicit numerical computation of the phase space trajectories
in the braneworld scenario, it is most convenient to rewrite the evolution
equations (\ref{FE}) and (\ref{EE}) for $H$ and $\phi$ as a set of two
first-order differential equations with two independent variables $\phi$ and
$\dot{\phi}$
\begin{eqnarray}
\frac{d\phi}{dt} & = & \dot{\phi}, \\
\frac{d\dot{\phi}}{dt} & = & -\frac{3\kappa_4}{\sqrt{6\sigma}}
\left[\frac{1}{2}\dot{\phi}^2+V(\phi)\right]\dot{\phi}-V'(\phi).
\end{eqnarray}

Let us consider two different examples. In the first example, we consider
a new inflation model in the braneworld scenario, with a potential
of Coleman-Weinberg type
\begin{equation}
V(\phi)=\lambda \phi^4 \left(\ln \frac{\phi^2}{\varphi^2}-\frac{1}{2}\right)
+\frac{1}{2}\lambda \varphi^4.
\end{equation}
We choose different initial conditions in the range
$\vert \phi_0 \vert < \varphi$ and
$\vert \dot{\phi}_0 \vert < \sqrt{\lambda}\varphi^2$, and we follow the evolution
until the field begins to oscillate around the true vacuum at $\pm \varphi$.
Figure 1 displays the trajectories in the $(x, y)$ plane for this model in the
braneworld scenario, where $x$ and $y$ are dimensionless coordinates
\begin{equation}
\phi=\varphi x,\ \dot{\phi}=\sqrt{\lambda}\varphi^2 y,\
t=\frac{1}{\sqrt{\lambda}\varphi}\eta,\ y=\frac{dx}{d\eta}\,. \nonumber
\end{equation}
We find that an initial kinetic term decays rapidly and it does not prevent
the onset of inflation. There is a curve that attracts most of the
trajectories. This curve corresponds to the slow rolling solution. Moreover,
comparing the phase portrait Figure 1 to Figure 2
reveals that the trajectories more rapidly reach
the slow rolling curve in the braneworld scenario than in the standard cosmology.
Assuming $\lambda \sim 1$, $\varphi \sim M_p$ and
$\varphi^4/ \sigma \sim 10$, the trajectory with $\dot{\phi}_0 \sim 0.3M_p^2$ reaches
the slow rolling curve at $\phi \sim 0.3M_p$ in Figure 1 while at
$\phi \sim 0.5M_p$ in Figure 2. 
\begin{figure}
\begin{center}
\includegraphics[angle=-90,scale=0.37]{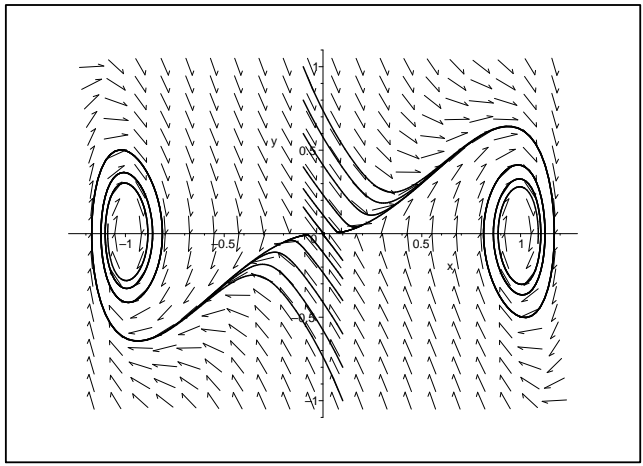}
\caption{Phase portrait for the scalar field with $V(\phi)=\lambda \phi^4
(\ln \frac{\phi^2}{\varphi^2}-\frac{1}{2}) +\frac{1}{2}\lambda \varphi^4$ in
rescaled coordinates $(x, y)$ in the braneworld scenario.}
\end{center}
\end{figure}
\begin{figure}
\begin{center}
\includegraphics[angle=-90,scale=0.37]{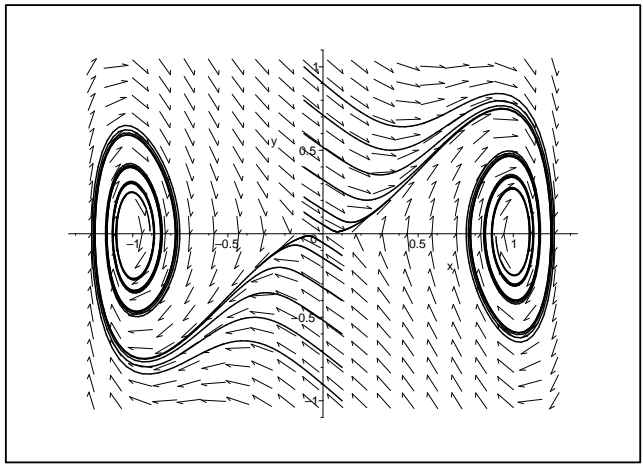}
\caption{Phase portrait for the scalar field with $V(\phi)=\lambda \phi^4
(\ln \frac{\phi^2}{\varphi^2}-\frac{1}{2}) +\frac{1}{2}\lambda \varphi^4$ in
rescaled coordinates $(x, y)$ in the standard FRW scenario.}
\end{center}
\end{figure}

In the second example, we investigate the simplest chaotic inflation model on
the brane driven by a scalar field with potential
\begin{equation}
V(\phi)=\frac{1}{2}m^2\phi^2.
\end{equation}
All solutions to these two equations can be represented as trajectories in the
two-dimensional phase space of $\phi$ and $\dot{\phi}$. Figure 3 displays
the trajectories in the $(x, y)$ plane for this model, where $x$ and
$y$ are dimensionless coordinates
\begin{equation}
\phi=mx,\ \dot{\phi}=m^2y,\ t=\frac{1}{m}\eta,\ y=\frac{dx}{d\eta}\,. \nonumber
\end{equation}
We see that there is a curve that attracts most of the trajectories. However, unlike
the new inflation case where only a small part of it corresponds to inflation,
in the chaotic case the attractor is the inflationary solution. Comparing Figure 3
to Figure 4, an initial kinetic term decays more rapidly in the braneworld scenario
than that in standard cosmology. The COBE constraints require $m \sim 10^{13}$
GeV~\cite{MWBH}. Assuming $V(\phi) / \sigma \sim 10$, the initial kinetic term
$\dot{\phi}_0 \sim 2\times 10^{31}$ GeV decays to zero at
$\phi \sim 3 \times 10^{17}$ GeV in Figure 3 while at
$\phi \sim 9 \times 10^{17}$ GeV in Figure 4.
\begin{figure}
\begin{center}
\includegraphics[angle=-90,scale=0.37]{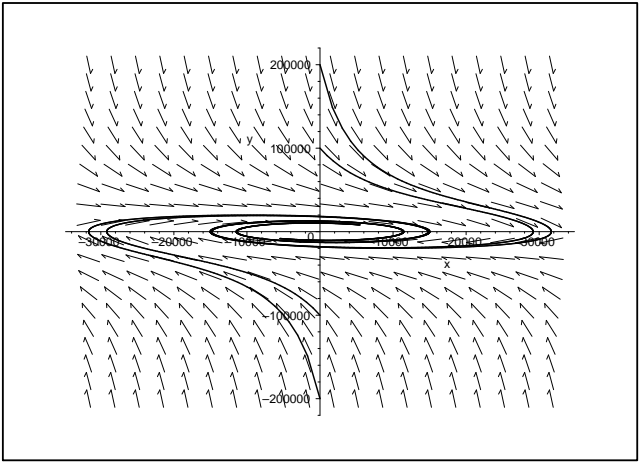}
\caption{Phase portrait for the scalar field with $V(\phi)=\frac{1}{2}m^2\phi^2$ in
rescaled coordinates $(x, y)$ in the braneworld scenario.}
\end{center}
\end{figure}
\begin{figure}
\begin{center}
\includegraphics[angle=-90,scale=0.37]{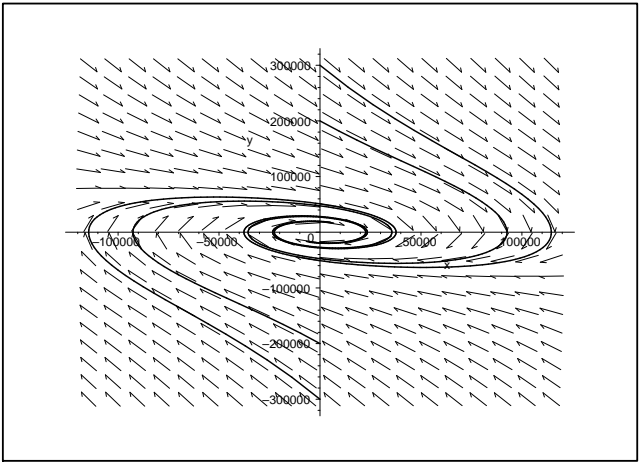}
\caption{Phase portrait for the scalar field with $V(\phi)=\frac{1}{2}m^2\phi^2$ in
rescaled coordinates $(x, y)$ in the standard FRW scenario.}
\end{center}
\end{figure}

\section{Attractor Behavior for Tachyonic Matter}

According to Sen~\cite{SEN}, the effective action of the tachyon field in the
Born-Infeld form can be written as
\begin{equation} \label{BIF}
S=\int d^4x\sqrt{-g}\left(\frac{1}{2\kappa_4 ^2}R
+V(\phi)\sqrt{1+\partial _\mu \phi \partial ^\mu \phi}\right),
\end{equation}
The rolling tachyon in a spatially flat FRW cosmological model can be
described by a fluid with a positive energy density $\rho$ and a negative
pressure $P$ given by
\begin{eqnarray}
\rho &=& \frac{V(\phi)}{\sqrt{1-\dot{\phi}^2}}, \\
P &=& -V(\phi)\sqrt{1-\dot{\phi}^2},
\end{eqnarray}
The evolution equation of the tachyon field minimally coupled to gravity and
the modified Friedmann constraint are
\begin{eqnarray}
\label{EET}
\frac{\ddot{\phi}}{1-\dot{\phi}^2}+3H\dot{\phi}+\frac{V'(\phi)}{V(\phi)}=0,\\
H^2=\frac{\kappa_4^2}{6\sigma} \frac{V^2(\phi)}{1-\dot{\phi}^2}.
\label{FET}
\end{eqnarray}
Differentiating Eq.(\ref{FET}) with respect to $t$ and substituting it in
Eq.(\ref{EET}) gives
\begin{equation}
\dot{H}=-3H^2\dot{\phi}^2.
\end{equation}
We may divide each side by $\dot{\phi}$ to eliminate the time dependence in
the Friedmann equation, obtaining the Hamilton-Jacobi equations
\begin{eqnarray}\label{HJT1}
\dot{\phi}=-\frac{H^\prime(\phi)}{3H^2(\phi)},\\
\frac{2\sigma}{3\kappa_4^2}\left[\frac{H^\prime(\phi)}{H(\phi)}\right]^2-
\frac{6\sigma}{\kappa_4^2}H^2(\phi)=-V^2(\phi).
\label{HJT2}
\end{eqnarray}

Once $H(\phi)$ has been specified, we can obtain the corresponding potential
and exact inflationary solution using (\ref{HJT1}) and (\ref{HJT2}). For example,
$H(\phi)=\phi^{-\alpha}$ gives the potential
\begin{equation}
V(\phi)=\frac{\sqrt{\sigma}}{\kappa_4} \left(6\phi^{-2\alpha}-\frac{2\alpha^2}{3}
\phi^{-2}\right)^{\frac{1}{2}}.
\end{equation}
The corresponding evolution of $\phi(t)$ is
\begin{equation}
\phi(t) = \left\{ \begin{array}{ll}
 \phi_0e^{\frac{2}{3}t} & \alpha = 2 \\
 \left(\phi_0^{2-\alpha}+\frac{(2-\alpha)\alpha}{3}t\right)^{\frac{1}{2-\alpha}} &
 \alpha \ne 2.
 \end{array} \right.
\end{equation}
Then we obtain the corresponding evolution of $a(t)$
\begin{equation}
a(t) \propto \left\{ \begin{array}{ll}
 \left(1+\frac{1}{3}\phi_0^{-1}t\right)^3 &  \alpha=1 \\
 \exp\left(-\frac{3}{4}\phi_0^{-2}e^{-\frac{4}{3}t}\right)& \alpha=2\\
 \exp\left[\frac{3}{2\alpha(1-\alpha)}\left(\phi_0^{2-\alpha}+
 \frac{2\alpha-\alpha^2}{3}t\right)^{\frac{2-2\alpha}{2-\alpha}}\right] & 
 \textrm{others}.
 \end{array} \right.
\end{equation}

Analogous to the preceding section we use the Hamilton-Jacobi formalism
to analysis the attractor behavior
of the tachyonic inflation in the braneworld scenario. Suppose $H_0(\phi)$ is
any solution to the full equation of motion, Eq.(\ref{HJT2}), which can be
either inflationary or non-inflationary. Consider a linear homogeneous perturbation 
$\delta H(\phi)$. It therefore obeys the linearized equation
\begin{equation}
H_0^\prime \delta H^\prime \simeq \left(18H_0^2-\frac{3\kappa_4^2}{2\sigma}
V^2\right)H_0\delta H,
\end{equation}
which has the general solution
\begin{equation}
\delta H(\phi)=\delta H(\phi_i)\exp \left[ \int_{\phi_i}^\phi \left( \frac{9H_0^3}
{H_0^\prime}+\frac{H_0^\prime}{H_0}\right) d\phi\right],
\end{equation}
We have used the fact that $H_0$ is any solution to Eq.(\ref{HJT2}). Since
$H^\prime _0$ and $d\phi$ have opposing signs, the integrand within the
exponential term is negative definite, and hence all linear perturbations
do indeed die away. That is, provided the potential is able to support inflation,
the inflationary solutions all rapidly approach one another, with exponential
rapidity once in the linear regime.

The tachyon potential $V(\phi) \to 0$ as $\phi \to \infty$, but its exact form is
not know at present. Sen has argued that the qualitative dynamics of string
tachyons can be described by (\ref{BIF}) with exponential potential~\cite{SEN2},
which is further investigated in support of this potential for the tachyon
system~\cite{SAMI}. The field equations (\ref{EET}) and (\ref{FET}) for tachyonic
matter on the brane with the exponential potential
\begin{equation}
V(\phi)=V_0e^{-\alpha \phi}
\end{equation}
can be cast in the form
\begin{eqnarray}
\frac{d\phi}{dt} & = & \dot{\phi}, \\
\frac{d\dot{\phi}}{dt} & = & (1-\dot{\phi}^2) \left[\alpha-\frac{3\kappa_4V(\phi)}
{\sqrt{6\sigma}}(1-\dot{\phi}^2)^{-\frac{1}{2}}\dot{\phi}\right].
\end{eqnarray}
We choose different initial conditions $\phi$ $(0 \leq \phi)$ and
$\dot{\phi}$ $(0 \leq \dot{\phi} \leq 1)$, and we follow the
evolution with the exponential potential numerically. Figure 5 displays
the trajectories in the $(\phi, \dot{\phi})$ plane for this model. We see that
there is a curve that attracts most of the trajectories. This curve corresponds
to the slow rolling solution. A sufficient tachyon inflation consistent
with the observational constraints in the braneworld scenario prefers the string
coupling $g_s \sim 10^{-16}$ and the string mass scale
$M_s \sim 10^{-7}M_p$ \cite{BBS}. Thus the string energy density
$V_0 \sim 10^{-15}M_p^4$. We find that the trajectory with
$\dot{\phi}_0 \sim 0.8 \times 10^{-7}M_p$ reaches the slow rolling solution
at $\phi \sim 0.7$ in Figure 5 in the braneworld scenario while at
$\phi \sim 1.2$ in Figure 6 in the standard cosmology, where we assume
$\alpha \sim 1/2$ and $V_0/ \sigma \sim 10$.
The numerical computation indicates that an initial kinetic term of tachyon field
decays more rapidly in the braneworld scenario than in the standard cosmology
if the brane tension remains sufficiently below the string energy density.
\begin{figure}
\begin{center}
\includegraphics[angle=-90,scale=0.37]{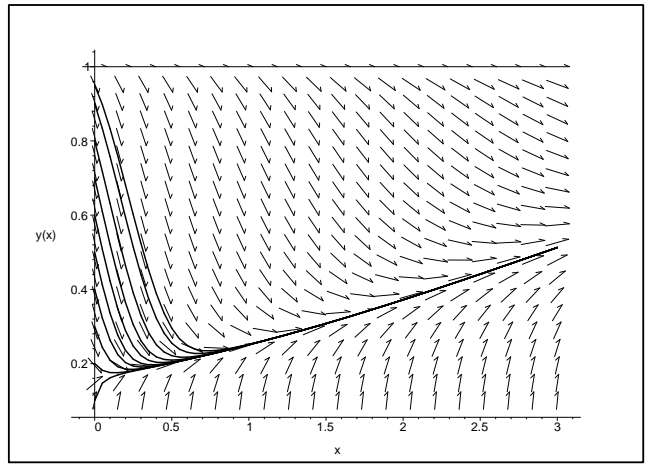}
\caption{Phase portrait for the tachyon field with $V(\phi)=V_0e^{-\alpha \phi}$
in rescaled coordinates $(\phi, \dot{\phi})$ in the braneworld scenario.}
\end{center}
\end{figure}
\begin{figure}
\begin{center}
\includegraphics[angle=-90,scale=0.37]{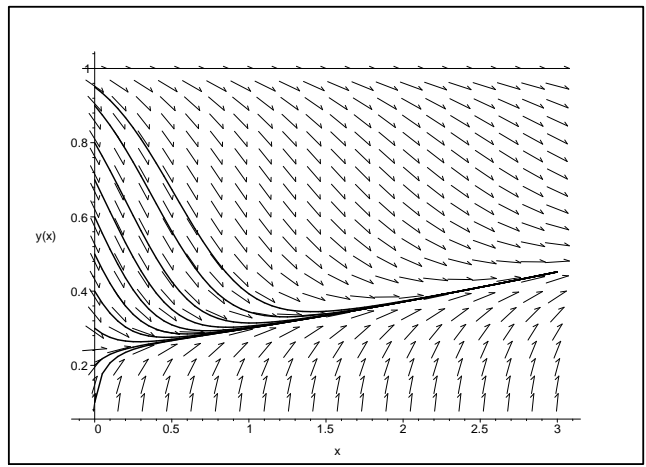}
\caption{Phase portrait for the tachyon field with $V(\phi)=V_0e^{-\alpha \phi}$
in rescaled coordinates $(\phi, \dot{\phi})$ in the standard FRW scenario.}
\end{center}
\end{figure}

The effective tachyon field action given by Eq.(\ref{BIF}) is a closed form
expression incorporating all the higher power of $\partial _\mu \phi$. It is
impossible to obtain a canonical kinetic term via a field redefinition. The
momentum  conjugate to $\phi$ is given by
\begin{equation}
\Pi=\frac{V(\phi)}{\sqrt{1-\dot{\phi^2}}}\dot{\phi}.
\end{equation}
Figure 7 and Figure 8 display the trajectories in the $(\phi, \Pi)$ plane for
this model in the braneworld scenario and in the standard FRW scenario
respectively, which are consistent with Figure 5 and Figure 6.
Comparing Figure 5 to Figure 7, when $\phi \to \infty$,
the kinetic term $\dot{\phi} \to 1$ and the momentum density $\Pi \to 0$
since $V(\phi) \to 0$. 
\begin{figure}
\begin{center}
\includegraphics[angle=-90,scale=0.37]{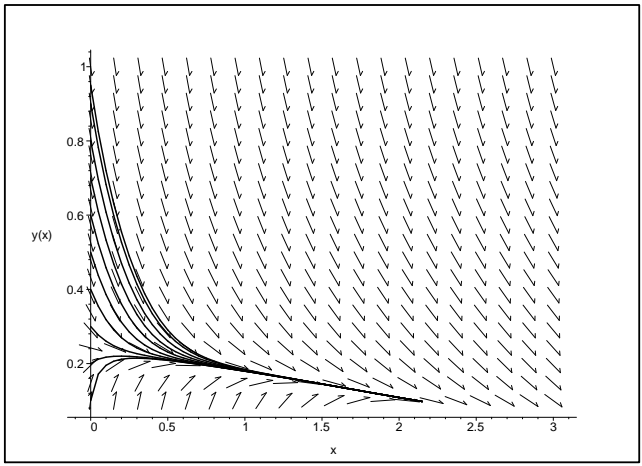}
\caption{Phase portrait for the tachyon field with $V(\phi)=V_0e^{-\alpha \phi}$
in rescaled coordinates $(\phi, \Pi)$ in the braneworld scenario.}
\end{center}
\end{figure}
\begin{figure}
\begin{center}
\includegraphics[angle=-90,scale=0.37]{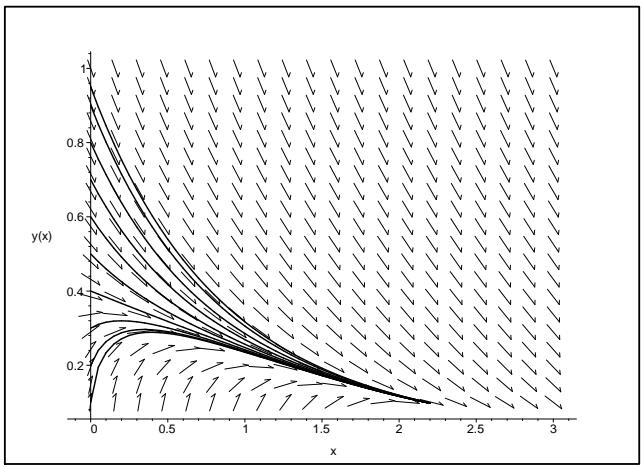}
\caption{Phase portrait for the tachyon field with $V(\phi)=V_0e^{-\alpha \phi}$
in rescaled coordinates $(\phi, \Pi)$ in the standard FRW scenario.}
\end{center}
\end{figure}

\section{Conclusions and Discussions}

We have derived the Hamilton-Jacobi equations of the inflation driven by a scalar
field or a tachyon field, which is confined on a three-dimensional brane in
five-dimensional Einstein gravity, and have obtained a set of exact inflationary solutions.
To demonstrate the attractor behavior, we use the Hamilton-Jacobi formalism, which
greatly simplifies the analysis. We restrict ourselves to linear homogeneous
perturbations, which is all that is needed because, classically at least, inflation does
indeed generate large smooth patches. If the perturbation is nonlinear, then the
solution is made more complicated but, because the full equation is only first order,
it is easy to see that solutions are compelled to approach one another regardless
of whether the perturbation is linear or not. For simplicity, we also assume that the
perturbations do not reverse the sign of $\dot{\phi}$, which can matter only if the
perturbation takes the field over the top of a maximum in the potential because
otherwise it will simply roll up, reverse its direction, and pass back down through
the same point, where it can be regarded as a perturbation on the original solution
with the same sign of $\dot{\phi}$.

Moreover, we have investigated two typical models for the scalar field: the new
inflation with the Coleman-Weinberg potential and the chaotic inflation with
the simplest potential, and have investigated the tachyon inflation on the brane
with the exponential potential. The phase space trajectories indicate that an
initial kinetic term decays rapidly and it does not prevent the onset of inflation,
that is, inflation on the brane possesses the attractor behaviour. In phase
space there is a curve that attracts most of the trajectories. This curve
corresponds to the slow rolling solution. Our numerical results show that
the initial kinetic terms more rapidly decay and so the trajectories more rapidly
reach the inflationary solution in the braneworld scenario than in the standard
cosmology. At low energies, $\rho \ll \sigma$, the modified Freidmann equation
(\ref{FRW}) reduces to the standard form since the term quadratic in the energy
density is negligible. However at high energies, $\rho \gg \sigma$, the term
quadratic in the energy density dominates the modified Freidmann equation. The
effect of the modified Friedmann equations (\ref{FE}) and (\ref{FET}) at high
energies is to increase the Hubble parameter by the new factor
$\rho / 2\sigma$. Thus the friction term $3H\dot{\phi}$ in the evolution
equation of the field (\ref{EE}) or (\ref{EET}) becomes larger, which leads to
more rapid decay of the initial kinetic term. Thus brane effects widen the
range of the initial conditions for successful inflation.

It is interesting to investigate cosmological evolutions in models where the effective
potential $V(\phi)=V_0+\frac{1}{2}m^2\phi^2$ may become negative for some values
of the field $\phi$ in the braneworld scenario. Phase portraits of such theories
in space of variables $(\phi,\dot{\phi},H)$ are different from phase portraits in the
standard usual cosmological scenario~\cite{FFKL}. For a flat universe $k=0$ with a
negative potential $V_0<0$, all trajectories in the three-dimensional phase space
$(\phi,\dot{\phi},H)$ are located at a paraboloid in contrast with a hyperboloid of
one sheet in the standard cosmological scenario. Trajectories both in the expanding
universe region $(H>0)$ and in the contracting universe region $(H<0)$ spiral in
towards the center in two-dimensional projection $(\phi,\dot{\phi})$ of the flat
universe hypersurface for this model. However, there appears a wormhole
connecting the expanding and contracting branches in the standard cosmological
scenario~\cite{FFKL}. In the braneworld scenario, cosmological evolution with
negative potential may have more fruitful phenomena, which is worth studying
further.

\section*{Acknowledgements}

It is a pleasure to acknowledge helpful discussions with Yun-Song Piao and
Rong-Gen Cai. This project was in part supported by NNSFC under Grant
No.10175070 and by NBRPC2003CB716300.

\end{document}